\RequirePackage{ifpdf}
\documentclass[12pt,letterpaper]{JHEP3}         
\usepackage{amssymb,amsfonts}

\usepackage{epsfig}

\def\be{\begin{eqnarray}}
\def\ee{\end{eqnarray}}
\newcommand{\nn}{\nonumber}
\newcommand\para{\paragraph{}}

\newcommand{\eqn}[1]{(\ref{#1})}

\def\Dslash{\,\,{\raise.15ex\hbox{/}\mkern-12mu D}}
\def\Dbarslash{\,\,{\raise.15ex\hbox{/}\mkern-12mu {\bar D}}}
\def\delslash{\,\,{\raise.15ex\hbox{/}\mkern-9mu \partial}}
\def\delbarslash{\,\,{\raise.15ex\hbox{/}\mkern-9mu {\bar\partial}}}
\def\pslash{\,\,{\raise.15ex\hbox{/}\mkern-9mu p}}
\def\calDslash{\,\,{\raise.15ex\hbox{/}\mkern-12mu {\cal D}}}

\def\lae{\mathrel{\mathop{\smash{\lower .5 ex \hbox{$\stackrel<\sim$}}}}}
\def\lae{\mathrel{\mathop{\smash{\lower .5 ex \hbox{$\stackrel>\sim$}}}}}

\title{Line Operators in the Standard Model}
\author{David Tong \\
Department of Applied Mathematics and Theoretical Physics, \\
University of Cambridge, UK\\
{\tt d.tong@damtp.cam.ac.uk}}

\abstract{There is an ambiguity in the gauge group of the Standard Model. The group is $G = SU(3) \times SU(2) \times U(1)/\Gamma$, where $\Gamma$ is a subgroup of ${\bf Z}_6$ which cannot be determined by current experiments. We describe how the electric, magnetic and dyonic line operators of the theory depend on the choice of $\Gamma$. We also explain how the periodicity of the theta angles, associated to each factor of $G$, differ.}

\begin{document}
\pagestyle{plain} \setcounter{page}{1}
\newcounter{bean}
\baselineskip16pt

\section{Introduction}

At the parochial distance scales accessible by experiment, the world in which we live is governed by the Standard Model. 
The gauge sector is one of the most beautiful constructs in theoretical physics, involving an intricate interplay between chiral fermions to ensure the cancellation of anomalies. Indeed, the matter content in one generation forms what is  arguably the simplest non-Abelian four-dimensional chiral gauge theory. 

\para
Despite the fact that the Standard Model is built around the idea of gauge symmetry, there is a little-advertised ambiguity in the choice of gauge group. We learn in kindergarten that we should take
\be \tilde{G} = U(1)\times SU(2) \times SU(3)\nn\ee
But this is not quite accurate. Experimental considerations tell us only that the gauge group is
\be G = {\tilde{G}}/{\Gamma}\nn\ee
where $\Gamma$ is a discrete group. As we review below, the matter content of the Standard Model is invariant under a suitably chosen  ${\bf Z}_6$ subgroup of $\tilde{G}$. For this reason, it is sometimes stated that one should  take the gauge group to include the quotient  $\Gamma = {\bf Z}_6$.  This, however, is putting the cart before the horse. At present, we can only say that the gauge group involves a quotient by $\Gamma$, which is a subgroup of ${\bf Z}_6$, i.e.
\be \Gamma = {\bf Z}_6,\ {\bf Z}_3,\ {\bf Z}_2\ {\rm or}\ {\bf 1}\nn\ee
Each of these possibilities defines a different theory and, ultimately, gives rise to different physics. The obvious questions are: which describes our world? And how can we tell?

\para
These are difficult questions to answer. Correlation functions of local operators in ${\bf R}^{1,3}$ depend only on the Lie algebra of the gauge group and are unaffected by global issues such as the choice of $\Gamma$. This means that no current experiment can distinguish between the four possibilities\footnote{The theoretical prejudice of unification suggests that $\Gamma = {\bf Z}_6$, since only then is $G$ is a subgroup of  $SU(5)$ or $Spin(10)$. It may well be true that this is the way Nature works. However, the  philosophy of this paper is to admit  our ignorance of the ultra-violet, and instead  use our knowledge of the infra-red to restrict what we may ultimately find as we explore higher energies.}. Nonetheless, the physics in flat space can depend in subtle ways on $\Gamma$ (and in more dramatic ways when spacetime has interesting topology). The purpose of this paper is to describe the crudest differences between the theories: the spectrum of line operators and the periodicities of theta angles.

\para
The fact that the spectrum of line operators depends on the global structure of the gauge group was emphasised by Aharony, Seiberg and Tachikawa \cite{ast}. The line operators can be thought of as heavy electric and magnetic test particles which can be used to probe the dynamics of the gauge fields.  Roughly speaking, taking a quotient restricts the allowed electric line operators but, in doing so, relaxes the constraint of Dirac quantisation and so frees up the allowed magnetic line operators. The first goal of this paper is to classify the allowed line operators for each choice of $\Gamma$.

\para
The authors of \cite{ast} also explained why  the periodicity of $\theta$-angles depends on the global structure of the gauge group. 
The second goal of this paper is to understand how the ranges of the various $\theta$-angles in the Standard Model are affected by the choice of $\Gamma$.  The $\theta$-angle for  $SU(3)$ is much discussed and the smallness of its (un)observed value, $\theta_3 \lesssim 10^{-10}$, is one of the great open problems in particle physics. In contrast there is seemingly no mystery about the $\theta$-angle for $SU(2)$ since the existence of the anomalous global $B+L$ symmetry means that it can always be rotated away.  (As we will see in Section 3, this argument needs a small correction.)

\para
Finally, there is also a  $\theta$-angle for $U(1)_Y$ hypercharge. This has received very little attention in the literature and with good reason, for Abelian gauge groups have no finite action configurations carrying topological charge. This means that, unlike their non-Abelian counterparts, the electric spectrum and correlation functions do not depend on $\theta_Y$. Nonetheless, Abelian $\theta$-angles can change the physics in more subtle ways. This happens, for example,  in the magnetic sector through the Witten effect \cite{witteneffect,frank}. It also happens when $\theta$-angles vary in space or, relatedly, in the presence of boundaries. Indeed, much of the rich and beautiful phenomenology of topological insulators can be understood as  a domain wall between $\theta_{em}=0$ and $\theta_{em}=\pi$ \cite{ti,mirage}.  We will see below that the periodicity of $\theta_Y$ is determined by $\Gamma$ and that, after electroweak symmetry breaking,  this affects the periodicity of the electromagnetic theta angle, which is a particular combination of $\theta_2$ and $\theta_Y$.

\para
The paper is organised as follows. 
Section \ref{reviewsec} is a review of the results of \cite{ast}, specifically how the line operators and associated theta angles differ for $SU(N)$ and $SU(N)/{\bf Z}_N$. Section 3 contains all the main results of the paper, including how the line operators and $\theta$-angles in the Standard Model depend on the choice of discrete quotient $\Gamma$. Particular attention is paid to the  values of $\theta$ for which the gauge sector is invariant under CP or, equivalently, under time reversal. We also describe the fate of the line operators under electroweak symmetry breaking, since this determines the allowed electromagnetic charges of particles.

\para
The distinctions between the different gauge groups $G=\tilde{G}/\Gamma$ described here are rather formal in nature. We end in Section 4 with some speculations on how these distinctions may manifest themselves physically.

\section{A Review of Line Operators}\label{reviewsec}

We start by reviewing the properties of line operators, described in \cite{ast},  that we will later need.

\para
Wilson lines are operators which describe the insertion of an infinitely massive, electrically charged particle sitting at the origin of space \cite{wilson}. They are labelled by representations $R$ of  the gauge group $G$, and given by
\be W_R = {\rm Tr}_R \,{\cal P}\exp\left(i\int dt\  A_0\right)\nn\ee
Wilson lines exist for all representations $R$, regardless of the dynamical matter content of the theory. This means, in particular, that different gauge groups $G$ will have a different spectrum of Wilson lines, even if they share a common Lie algebra ${\bf g}$. We denote the weight lattice of ${\bf g}$ as $\Lambda_w$. Then the representations of the universal cover of $G$ --- which we denote as $\tilde{G}$ ---  are labelled by points in the lattice $\Lambda_w/W$ where $W$ is the Weyl group. Representations of $G = \tilde{G}/\Gamma$ are labelled by an appropriate sublattice $\Lambda_w/W$.

\para
't Hooft lines are operators which describe the insertion of an infinitely massive, magnetically charged particle sitting at the origin of space \cite{thooft}.  They are best thought of as defect operators, in which the gauge fields have suitable boundary conditions imposed on the worldline of the magnetic source \cite{kapustin}.  For a gauge group $G$, 't Hooft lines are labelled by some sublattice of $\Lambda_{mw}/W$ where $\Lambda_{mw}$ is the magnetic weight lattice \cite{gno}; it is the weight lattice of the dual Lie algebra ${\bf g}^\star$, or the dual of the root lattice of ${\bf g}$.

\para
More generally, we may probe the theory with a dyonic Wilson-'t Hooft line. These are labelled by some sublattice of $(\Lambda_w\times \Lambda_{mw})/W$ \cite{kapustin}. We will denote the weights as $(\lambda^e,\lambda^m)\in \Lambda_w\times \Lambda_{mw}$, with the identification $(\lambda^e,\lambda^m)\sim (w\lambda^e,w\lambda^m)$, where $w\in W$.

\para
Given a choice of gauge group $G= \tilde{G}/\Gamma$, we would like to determine the allowed spectrum of  line operators. This  problem was solved in \cite{ast} for connected  groups. (Here we consider only line operators which enjoy an independent existence, as opposed to  those which survive only on the boundary of a surface operator.) Certain line operators exist regardless of the choice of the discrete quotient  $\Gamma$. For the Wilson lines, these correspond to the root lattice $\Lambda_r\subset \Lambda_w$. They include, for example, the Wilson line in the adjoint representation.  Different choices of $\Gamma$ then determine which of the remaining Wilson lines are allowed. These can be characterised by their transformation under $\Lambda_w/\Lambda_r = Z(\tilde{G})$, the centre of $\tilde{G}$. 

\para
A similar story plays out for the 't Hooft lines: for every choice of $G = \tilde{G}/\Gamma$ there is a 't Hooft line corresponding to the  co-root lattice $\Lambda_{cr}\subset \Lambda_{mw}$ or, equivalently the root lattice of ${\bf g}^\star$. The remaining 't Hooft lines are characterised by  $\Lambda_{mw}/\Lambda_{cr} = Z(\tilde{G})$ and their admissibility depends on the choice of $\Gamma$.

\para
The upshot of this is that line operators  decompose into classes, labelled by the 
\be
(z^e,z^m) \in Z(\tilde{G}) \times Z(\tilde{G})\nn\ee
Furthermore, the line operators form a group; the operator product gives rise to an addition law while the reversal of orientation provides an inverse. These properties are enough to ensure that if one member of the class labelled by $(z^e,z^m)$ is present in the theory the the entire class is present. 

\begin{figure}[htb]
\begin{center}
\epsfxsize=6.2in\leavevmode\epsfbox{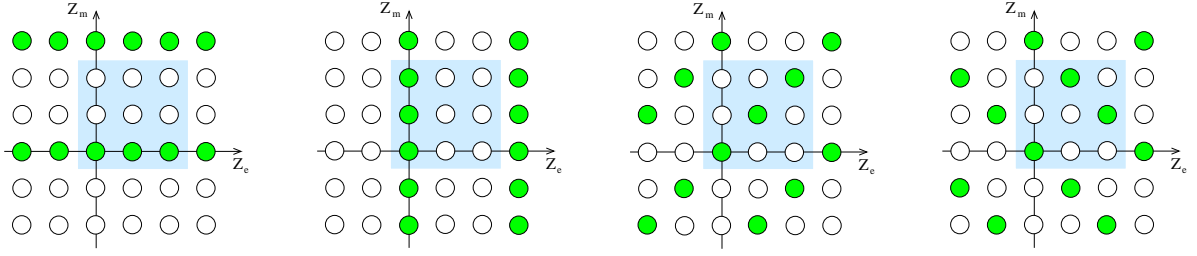}
\end{center}
\caption{The spectrum of line operators for: a) $SU(3)$;  b) $SU(3)/{\bf Z}_3$ at $\theta=0$;  c) $SU(3)/{\bf Z}_3$ at $\theta=2\pi$;  d) $SU(3)/{\bf Z}_3$ at $\theta=4\pi$.}
\label{su3lines}
\end{figure}

\para
The admissible electric line operators $(z^e,0)$ are determined by the choice of gauge group $G=\tilde{G}/\Gamma$, with only those $z^e$ which are invariant under $\Gamma$ present. The magnetic and dyonic line operators are then restricted by Dirac quantisation. To be specific, consider gauge groups with centre $Z(\tilde{G}) = {\bf Z}_N$. In this case both $z^e$ and $z^m$ are integers mod $N$. Two lines $(z^e,z^m)$ and $(z^{\prime\,e},z^{\prime\,m})$ can both exist as operators only if
\be z^ez^{\prime\,m} - z^mz^{\prime\,e} = 0\ \ {\rm mod}\ \ N\label{diracq}\ee
This is the Dirac quantisation condition for non-Abelian lines. The important role played by the centre of the gauge group in Dirac quantisation was first pointed out by Corrigan and Olive \cite{corrigan}.

\subsubsection*{An Example: $SU(N)$ vs $SU(N)/{\bf Z}_N$}

For $g=su(N)$, the centre of the group is $Z(SU(N)) = {\bf Z}_N$.  When the gauge group is $G=S(N)$, all electric line operators are allowed; that is $z^e=0,1,2,\ldots, N-1$. In contrast, the magnetic line operators are restricted to lie on the magnetic root lattice, meaning that $z^m=0$. The resulting set of line operators for $SU(3)$  are shown in Figure \ref{su3lines}a, plotted in the $(z^e,z^m)$ lattice. The figure shows the charges extended to $z^{e,m}\in {\bf Z}$;  the key physics can be seen in the grey box, which restricts  to $z^{e,m} =0,1,2$. The lattice is then formed by tiling this box.

\para
For $SU(N)/{\bf Z}_N$, the only purely electric line operators lie on the root lattice. They have $(z^e,z^m)=(0,0)$. The magnetic line operators are more interesting and there are several, different solutions to the quantisation condition \eqn{diracq}. The simplest such solution is to admit the purely magnetic line operators $(0,z^m)$ with $z^m=0,1,2,\ldots N-1$. No further dyonic operators are then allowed. The resulting spectrum of line operators for $G=SU(3)/{\bf Z}_3$ is shown in Figure \ref{su3lines}b.

\para
For $G=SU(N)/{\bf Z}_N$ (but not for all other gauge groups \cite{ast}), other solutions to \eqn{diracq} are generated by the theta angle. This arises through the Witten effect \cite{witteneffect} -- which holds for line operators as well as dynamical particles \cite{kapustin,henningson} -- and ensures that, in the presence of $\theta\neq 0$,  purely magnetic line operators turn into dyonic line operators. In particular, as $\theta \rightarrow \theta+2\pi$, each line operator picks up an electric charge given by
\be \lambda^e \rightarrow \lambda^e + \lambda^m\ \ \ \Rightarrow\ \ \ z^e\rightarrow z^e+z^m\nn\ee
For the $G=SU(N)$ theory, the set of line operators transforms into itself under $\theta\rightarrow \theta +2\pi$. In contrast, for $G=SU(N)/{\bf Z}_N$, the set of line operators turns into something different. This means that the theory with $\theta=0$ is not the same as the theory with $\theta=2\pi$. Instead, for $G=SU(N)/{\bf Z}_N$, the theta angle takes values in the range
\be \theta \in [0,2\pi N)\nn\ee
This extended range of $\theta$  is associated to the fact that $G=SU(N)/{\bf Z}_N$ admits instantons with fractional charge $1/N$. Starting with the set of line operators generated by $(z^e,z^m) = (0,1)$, under a shift $\theta\rightarrow \theta + 2\pi k$ we have a new theory with line operators generated by $(z^e,z^m)=(k,1)$. For $G=SU(3)/{\bf Z}_3$,  the set of line operators are shown in  Figures \ref{su3lines}c and \ref{su3lines}d for $k=1$ and $k=2$ respectively. 

\subsubsection*{How Does the Dynamics Differ?}

In four-dimensional Minkowski spacetime, ${\bf R}^{3,1}$, the difference between Yang-Mills with gauge group $SU(N)$ and  $SU(N)/{\bf Z}_N$ is rather formal. In particular, any local observer is blind to the distinction. Nonetheless, the different line operators mean that there are subtle differences between the two theories. This appears, for example, after confinement. In either theory, the confining phase  can be viewed as arising through the condensation of magnetic monopoles with charge $\lambda^m\in \Lambda_{mr}$. For $SU(N)$ Yang-Mills, these are the minimally charged monopoles. However, for $SU(N)/{\bf Z}_N$, these are not the minimum charge. This means that (at $\theta=0$) this theory exhibits topological order, with an emergent {\it magnetic} discrete ${\bf Z}_N$ gauge symmetry in the infra-red \cite{ast}.

\para
The difference between the two theories takes on a more meaningful role when the theory is compactified on a space with non-trivial topology, since now the Wilson line for the ${\bf Z}_N$ gauge symmetry can get an expectation value, resulting in different physics in the lower dimension.  This is perhaps clearest in the ${\cal N}=1$ super Yang-Mills, where one has more control over the dynamics. When compactified on a circle, or on a higher dimensional torus,   the Witten indices  for $SU(N)$ and $SU(N)/{\bf Z}_N$ differ \cite{witten,yuji}.  

\section{Line Operators in the Standard Model}

In this section, we extend the analysis of \cite{ast} to the non-connected gauge group
\be G  = \frac{U(1)_{\tilde{Y}} \times SU(2) \times SU(3)}{\Gamma}\nn\ee
where $\Gamma \subseteq {\bf Z}_6$. The quotient group $\Gamma$ lies in the centres of $SU(2)$ and $SU(3)$, combined with a suitable $U(1)_{\tilde{Y}}$ rotation. The quotient $\Gamma={\bf Z}_6$ is generated by
\be \xi = e^{2\pi i q/6} \otimes \eta \otimes \omega\nn\ee
where $\eta\in Z(SU(2))$ obeys $\eta^2=1$ and $\omega\in Z(SU(3))$ obeys $\omega^3=1$ and  $q$ is the $U(1)_{\tilde{Y}}$ charge. The quotient $\Gamma ={\bf Z}_3$ is generated by $\xi^2$ and the quotient $\Gamma = {\bf Z}_2$ is generated by $\xi^3$.

\para
Line operators are now labelled by three electric charges and three magnetic charges, one pair for each factor of the gauge group. As reviewed in Section \ref{reviewsec}, for non-Abelian gauge groups the line operators fall into classes, labelled by $z_2^e,z_2^m=0,1$ for $SU(2)$ and $z_3^e,z_3^m=0,1,2$ for $SU(3)$. We also require the additional labels $(q,g)$ to describe the electric and magnetic charge under $U(1)_{\tilde{Y}}$. We chose conventions\footnote{The convention that charges under $U(1)_{\tilde{Y}}$ are integer valued is standard in more formal areas of field theory, but differs from the usual normalisation of hypercharge in the Standard Model, which is given by $Y = \tilde{Y}/6$. Later, we will make the assumption that the minimal hypercharge $q=1$ is realised in the Standard Model (by the left-handed quarks). } such that $q\in {\bf Z}$   and, in the {\it absence} of any discrete quotient, $g\in {\bf Z}$ as well. However, as we will see, the presence of a discrete quotient $\Gamma \neq {\bf 1}$ means that $g$ can take fractional values.

\para
The Dirac quantisation condition is simplest to state between purely electric and purely magnetic lines: it is
\be (-1)^{z_2^ez_2^m} (e^{2\pi i/3})^{z_3^ez_3^m} e^{-2\pi i qg} = 1\nn\ee
Or, equivalently,
\be 3z_2^ez_2^m + 2 z_3^ez_3^m - 6qg \in 6{\bf Z}\label{this}\ee
We deal with each choice of $\Gamma ={\bf 1}$, ${\bf Z}_2$, ${\bf Z}_3$ and ${\bf Z}_6$ in turn. We start by describing the spectrum of line operators when  $\theta=0$ for each factor of the gauge group; we will subsequently see how the spectrum changes with $\theta$.

\para
\underline{$\Gamma= {\bf 1}$:}\ With no quotient, there is no restriction on the allowed electric line operators: the theory contains Wilson lines with charges $z^e=0,1$ and  $z_3^e=0,1,2$, dressed with any Abelian $q\in {\bf Z}$. Solutions to \eqn{this} then require $z_2^m=0$ mod 2 and $z_3^m=0$ mod 3 while $g\in {\bf Z}$.

\DOUBLEFIGURE{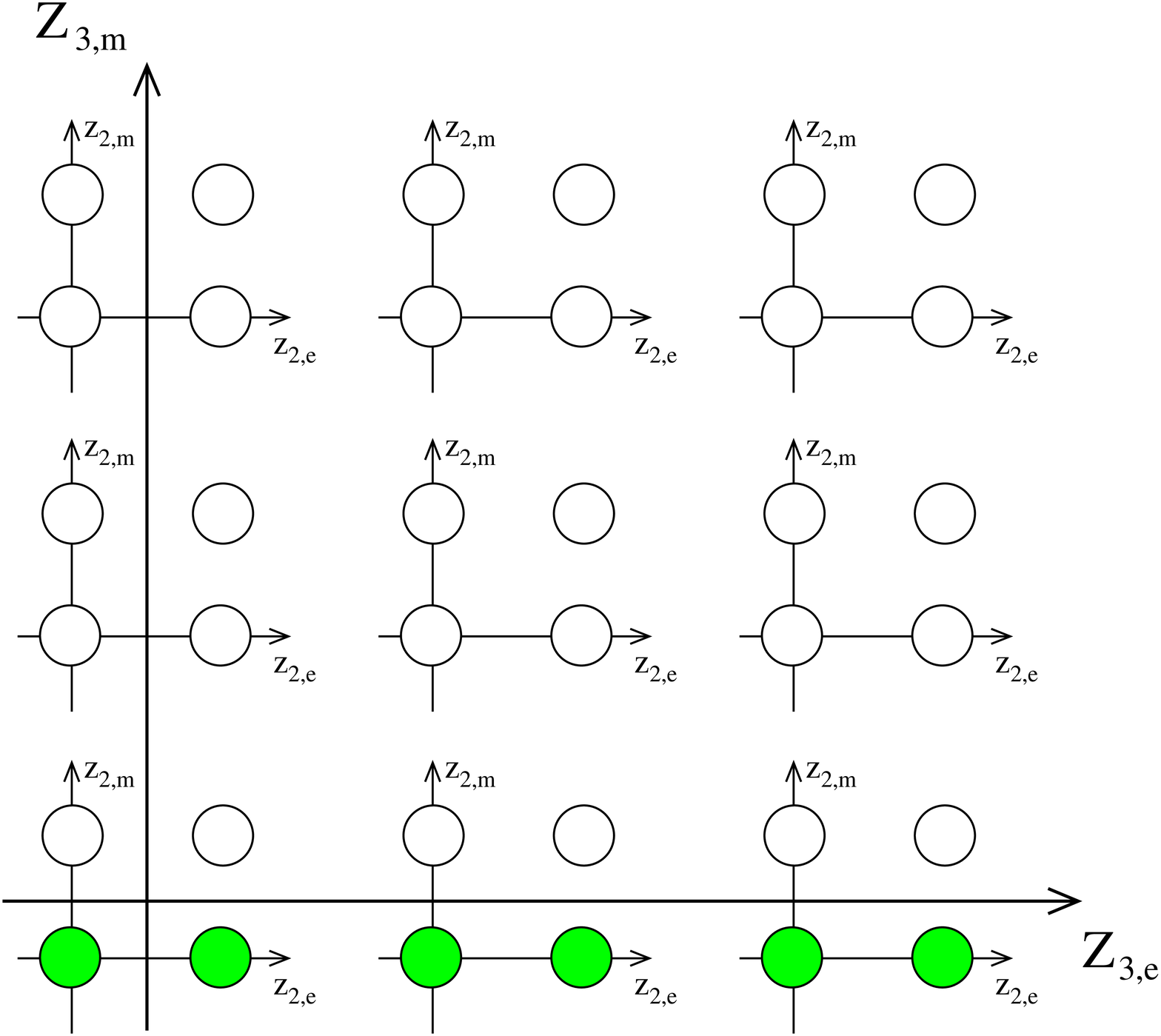,width=200pt}{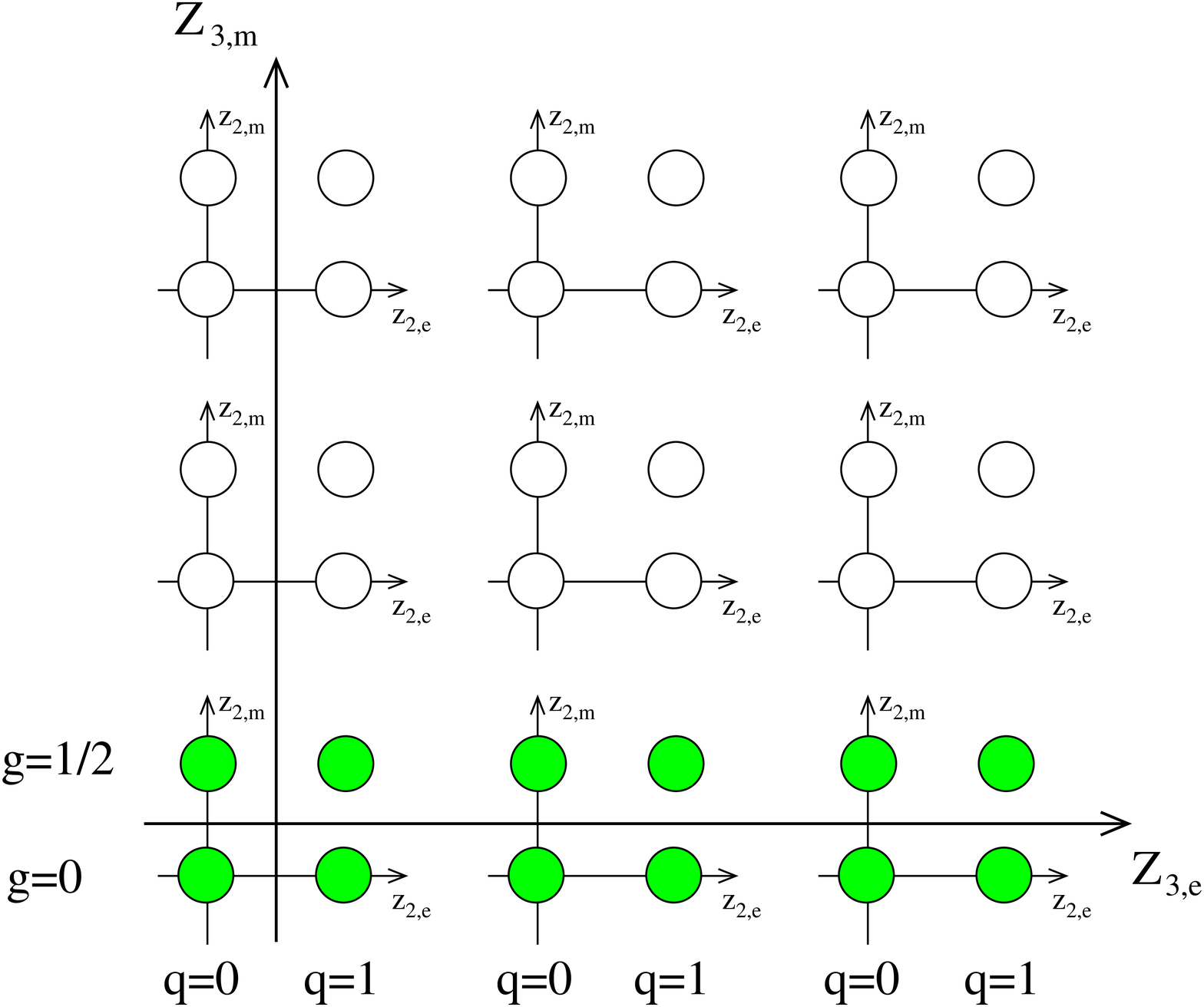,width=200pt}{$\Gamma={\bf 1}$. The Abelian lines are generated by $(q,g)=(1,0)$ and $(0,1)$}{$\Gamma={\bf Z}_2$. The Abelian lines are generated by $(q,g)=(2,0)$ and $(0,1)$}

\para
The resulting electric and magnetic charges of line operators are depicted in Figure 2. This, and subsequent figures, show the four possible $(z_2^e,z_2^m)$ charges superposed on the nine possible $(z_3^e,z_3^m)$ charges. In this case, we can add Abelian line operators with $(q,g)=(1,0)$ and $(q,g)=(0,1)$. 

\para
\underline{$\Gamma= {\bf Z}_2$:} Wilson lines must now be invariant under $\xi^3$ This means that electric lines with $z_2^e=0$ must have $q$ even, while those with $z_2^e=1$ must have $q$ odd. Each of these can have any $z_3^e=0,1,2$.

\para
The quantization condition condition \eqn{this} still requires $z_3^m=0$ mod 3. However, now magnetic lines exist with $z_2^m=1$ provided they are accompanied by Abelian magnetic charge $g=\frac{1}{2}$. We can add to these Abelian line operators with $(q,g)=(2,0)$ and $(q,g)=(0,1)$. The resulting spectrum of line operators is that of $U(2) \times SU(3)$ and is shown in Figure 3.

\para
\underline{$\Gamma= {\bf Z}_3$:} Wilson lines must now be invariant under $\xi^2$. This mean that electric lines must have $q=z_3^e$ mod 3. 
 Each of these can have any $z_2^e=0,1$.

 \para
 The quantization condition condition \eqn{this} now allows lines with $SU(3)$ magnetic charge $z_3^m=0,1,2$, as long as they are accompanied by Abelian magnetic charge $g=z_3^m/3$. No $SU(2)$ magnetic charge is allowed: $z_2^m=0$ mod 2. We can add to these Abelian line operators with $(q,g)=(3,0)$ and $(q,g)=(0,1)$. The resulting spectrum of line operators is that of $SU(2) \times U(3)$ and is shown in Figure 4. 

 \DOUBLEFIGURE{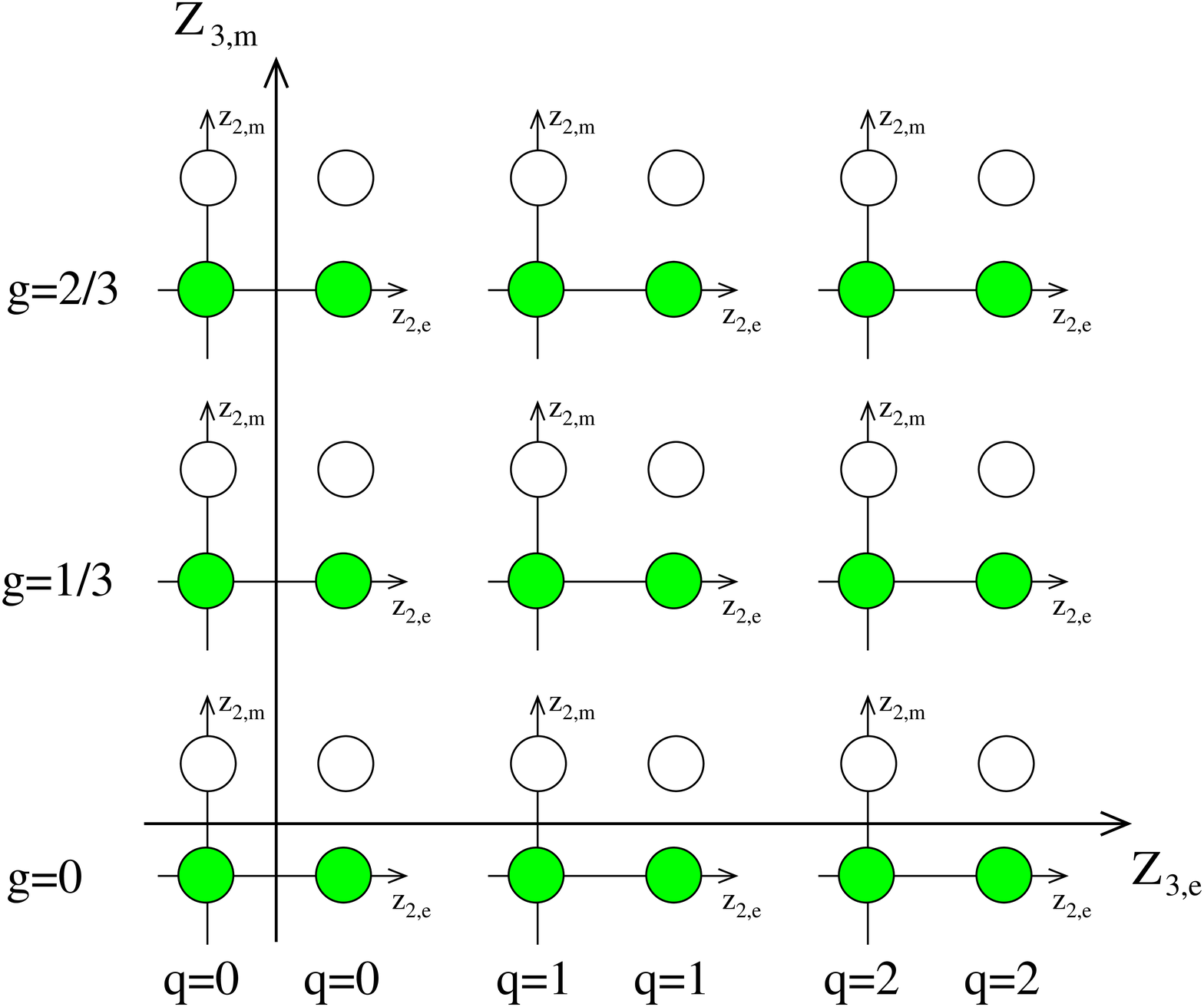,width=200pt}{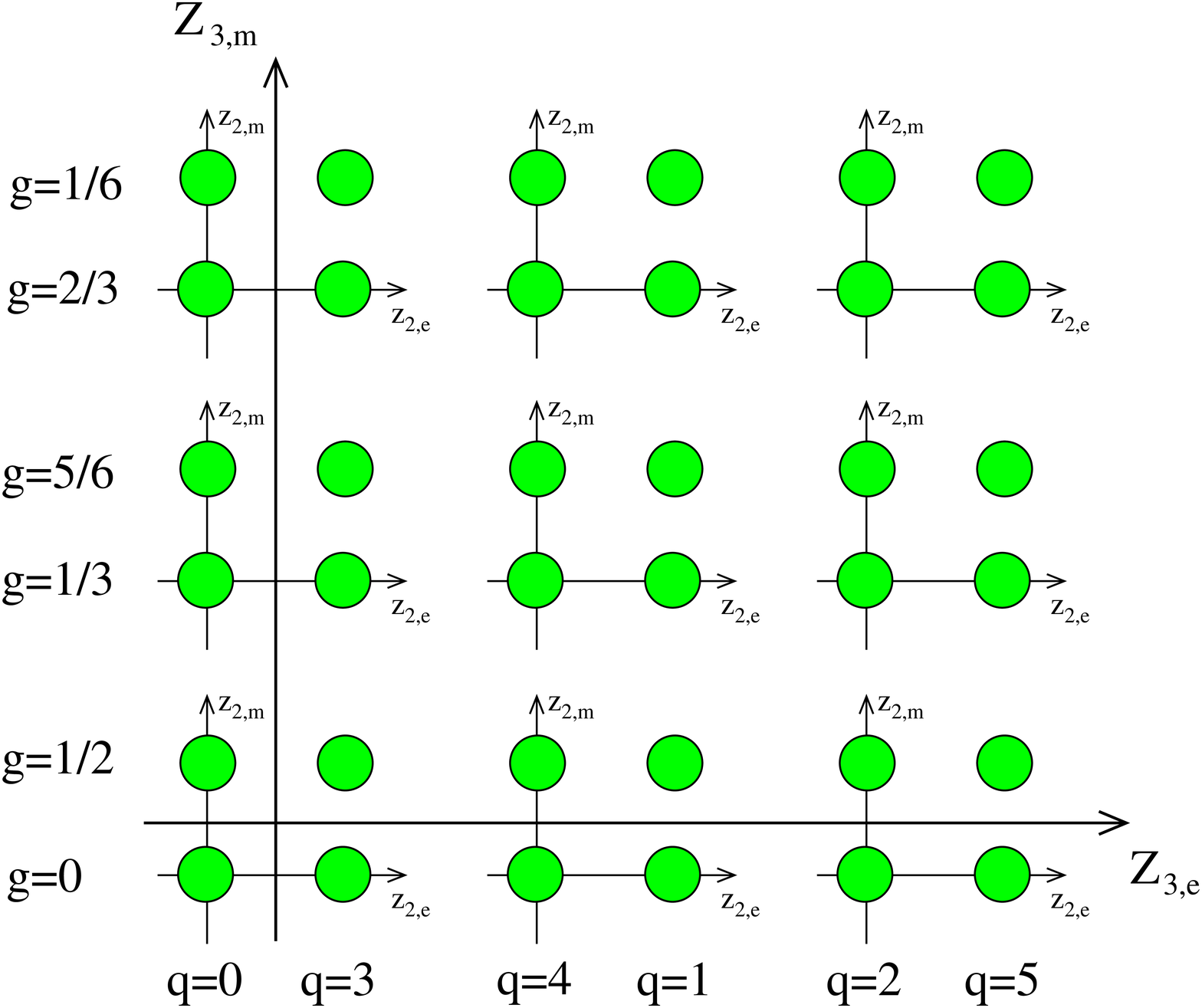,width=200pt}{$\Gamma={\bf Z}_3$. The Abelian lines are generated by $(q,g)=(3,0)$ and $(0,1)$}{$\Gamma={\bf Z}_6$. The Abelian lines are generated by $(q,g)=(6,0)$ and $(0,1)$}\label{b}

\para
\underline{$\Gamma= {\bf Z}_6$:}  Purely electric operators are invariant under $\xi$. This means that the Abelian charge is $q = 3z_2^e - 2z_3^e$ mod 6. It is noticeable that all  fundamental fermions in the Standard Model obey this relationship between the charges\footnote{Wilson lines should be thought of as the insertion of infinitely heavy particles, and so are not directly associated to  these massless, chiral fermions. Nonetheless, both sit in representations of the gauge group with $\Gamma = {\bf Z}_6$.}. Specifically, the representations under $U(1)\times SU(2)\times SU(3)$ are
\be &\mbox{Leptons:}&\ \   l_L: \ \ ({\bf 2},{\bf 1})_{-3} \ \ \ \Rightarrow \ \ \ (z_2^e,z_3^e)_q = (1,0)_{-3} \nn\\ 
&&\ \  e_R:\ \ ({\bf 1},{\bf 1})_{-6}\ \ \, \Rightarrow\ \ \   (z_2^e,z_3^e)_q = (0,0)_{-6} \nn\\ 
&\mbox{Quarks:}&\ \  q_L: \ \ ({\bf 2},{\bf 3})_{+1} \ \ \ \Rightarrow \ \ \ (z_2^e,z_3^e)_q = (1,1)_{+1} \nn\\ 
&&\ \  u_R:\ \ ({\bf 1},{\bf 3})_{+4}  \ \ \ \Rightarrow \ \ \ (z_2^e,z_3^e)_q = (0,1)_{+4} \nn\\ 
&&\ \  d_R:\ \ ({\bf 1},{\bf 3})_{-2}  \ \ \ \Rightarrow \ \ \ (z_2^e,z_3^e)_q = (0,1)_{-2}\nn\ee
We could add to this  the right-handed neutrino $\nu_R$ which is a gauge singlet.
The Higgs also obeys the relationship between electric charges, sitting in the representation $({\bf 2},{\bf 1})_{3}  \ \Rightarrow \  (z_2^e,z_3^e)_q = (1,0)_{3}$. The fact that all Standard Model fields satisfy $q = 3z_2^e - 2z_3^e$ mod 6 is, of course, equivalent to the statement made in the introduction that the Standard Model gauge group is consistent with $U(1)\times SU(2)\times SU(3)/{\bf Z}_6$.

\para
The  quotient $\Gamma = {\bf Z}_6$ allows for the richest spectrum of magnetic line operators. Now purely magnetic operators exist for any choice of $(z_2^m,z_3^m)$ provided they are accompanied by an Abelian magnetic charge $6g = 3z_2^m + 2z_3^m$ mod 6. For example, a basis of magnetic operators is $(z_2^m,z_3^m)_g =(1,0)_{1/2}$ and $(0,1)_{1/3}$. We can add to these Abelian line operators with $(q,g) = (6,0)$ and $(0,1)$. The resulting spectrum is shown in Figure 5.

\subsection{Theta Angles}

We can now ask how the spectrum of line operators changes as we vary the $\theta$-angles. A priori, there are three such angles, one for each factor of the gauge group. We call these $\theta_{\tilde{Y}}$, $\theta_2$ and $\theta_3$. Within the framework discussed in this paper it makes sense to ask how the line operators vary under each of these. 

\para
Before we proceed, it is worth reviewing the role of theta angles in the Standard Model.  The most discussed is the QCD theta angle, $\theta_3$. Bounds on strong CP violation restrict $\theta_3\lesssim 10^{-10}$. (The value $\theta_3=\pi$ also preserves CP but the meson spectrum derived from the chiral Lagrangian differs from the observed values \cite{baluni,crewther}.)

\para
It is usually stated that the  weak theta angle, $\theta_2$, can be rotated away in the  Standard Model . This follows from the existence of an anomalous global symmetry $B+L \rightarrow {\bf Z}_{N_f}$ where $N_f=3$ is the number of generations. 
We will revisit this below. 

\para
Finally, there is very little, if any, discussion of the theta angle for hypercharge $\theta_{\tilde{Y}}$. This changes neither the spectrum nor correlation functions of local operators. Nonetheless, it can play a role in the presence of magnetic monopoles or boundaries of space.  Correspondingly, it also changes the spectrum of line operators.

\para
Here we start by ignoring the effects of global anomalies and focus on the spectrum of line operators and the Witten effect. 
As reviewed in Section \ref{reviewsec}, for simple gauge groups, a quotient by the centre has the effect of extending the range of $\theta$. In the present context, the quotient $\Gamma = {\bf Z}_p$ extends the range of the Abelian theta angle only\footnote{For the hypercharge with normalisation $Y=\tilde{Y}/6$, the range is $36 \theta_Y\in [0,2\pi  p^2)$, so $\theta_Y\in [0,2\pi)$ when $\Gamma={\bf Z}_6$.}. We have
\be \theta_2,\theta_3\in [0,2\pi)\ \ \ {\rm and}\ \ \ \theta_{\tilde{Y}} \in [0,2\pi p^2)\label{thetarange}\ee
This is simplest to see for the case of ${\bf Z}_2$ and ${\bf Z}_3$ where the gauge group is   $G=U(2)\times SU(3)$ and $G=SU(2)\times U(3)$ respectively. Here
\be U(N) = \frac{U(1)\times SU(N)}{{\bf Z}_N}\nn\ee
We denote the $U(1)$ gauge field as $\tilde{a}$, the $SU(N)$ gauge field as $a$ and their corresponding field strengths as $\tilde{f}$ and $f$. The theta terms for the $U(1)\times SU(N)$ theory are
\be S_\theta =  \frac{\theta_N}{16\pi^2}\int\ {\rm tr}\, (f\, {}^\star\!f)  + \frac{\tilde{\theta}}{16\pi^2}\int \  \tilde{f}\, {}^\star\!{\tilde f} \nn\ee
To describe the $U(N)$ theory, we introduce the canonically normalised gauge field
\be b = a + \tilde{a}{\bf 1}_N\nn\ee
with corresponding field strength $g$. The theta terms can then be written as
\be S_\theta = \frac{\theta_N}{16\pi^2}\int \ {\rm tr}\, (g\,{}^\star\!g)  + \frac{\tilde{\theta} - N\theta_N}{16\pi^2 N^2}\int \ ({\rm tr}\, g)\, {}^\star({\rm tr}\, g)\label{untheta}\ee
We see that $\theta_N\in [0,2\pi)$ while $\tilde{\theta}\in [0,2\pi N^2)$. 

\para
Similarly, for the ${\bf Z}_6$ quotient, one can check that the spectrum of line operators is invariant under the identification \eqn{thetarange}, with $\tilde{\theta} \in [0,72\pi)$.

\para
(Some examples with the ${\bf Z}_6$ quotient: the theory with $\theta_2=0$, $\theta_3=2\pi$ and Abelian theta angle $\tilde{\theta}$ has the same spectrum of line operators as the theory with $\theta_2=\theta_3=0$ and Abelian theta angle $\tilde{\theta}'=\tilde{\theta} + 48\pi$; the theory with $\theta_2=2\pi$, $\theta_3=0$ and Abelian theta angle $\tilde{\theta}$ has the same spectrum of line operators as the theory with $\theta_2=\theta_3=0$ and Abelian theta angle $\tilde{\theta}'=\tilde{\theta} + 36\pi$.)

\subsection*{Time Reversal Invariant States}

It is interesting to ask: for which values of the $\theta$ angles does the theory respect time reversal invariance or, equivalently, CP? There are eight possibilities, two for each factor of the gauge group. 

\para
\underline{$\Gamma= {\bf 1}$:}\ \ All theta angles have periodicity $2\pi$. It is well known that $\theta_2,\theta_3$ and $\tilde{\theta}$ can each take values 0 or $\pi$. 

\para
\underline{$\Gamma= {\bf Z}_2$:}\ \ Here $\tilde{\theta}\in [0,8\pi)$. The values $\theta_3=0$ and $\pi$ are both time reversal invariant. We can read off the $U(2)$ $\theta$ angles from \eqn{untheta}. When $\theta_2=0$, $\tilde{\theta}=0$ or $4\pi$; when $\theta_2=\pi$ then $\tilde{\theta} = 2\pi$ or $6\pi$.

\para
\underline{$\Gamma= {\bf Z}_3$:}\ \ Here $\tilde{\theta} \in [0,18\pi)$. The values $\theta_2=0$ and $\pi$ are both time reversal invariant. The $U(3)$ theta angles can be  $\theta_3=0$ and $\tilde{\theta}  = 0$ or $9\pi$. Alternatively, we can have  $\theta_3=\pi$ and $\tilde{\theta} = 3\pi$ or $12\pi$. 

\para
\underline{$\Gamma= {\bf Z}_6$:}\ \ We have $\tilde{\theta}\in [0,72\pi)$. The time reversal invariant theories have
\begin{itemize}
\item $\theta_2=0$ and $\theta_3=0$ and $\tilde{\theta} = 0$ or $36\pi$.
\item $\theta_2=0$ and $\theta_3=\pi$ and $\tilde{\theta} = 12\pi$ or $48\pi$.
\item $\theta_2=\pi$ and $\theta_3=0$ and $\tilde{\theta} = 18\pi$ or $54\pi$.
\item $\theta_2=\pi$ and $\theta_3=\pi$ and $\tilde{\theta} =30\pi$ or $66\pi$.
\end{itemize}

\subsection*{The Effect of Global Anomalies}

The chiral nature of the Standard Model means that the theta angle $\theta_2$ can be rotated away. Here we review this argument. In fact, as we will see,  a more careful statement is that a linear combination of $\theta_2$ and $\tilde{\theta}$ can be removed.

\para
In general, theta angles can be rotated away if the theory admits a  continuous global symmetry which suffers a mixed anomaly with the gauge symmetry. This arises most naturally in the presence of a massless chiral fermion. (For example, a massless up quark provides an elegant solution to the strong CP problem, albeit one that appears not to be favoured by Nature). However, even with non-vanishing Yukawa couplings, so that all fermions have a mass, the Standard Model still admits two global symmetries: these are lepton and baryon number:

%
\begin{center}
\begin{tabular}{c|cccccc}
 & $l_L$ &  $q_L$ & $e_R$ & $u_R$ & $d_R$ & $\nu_R$  \\ \hline
L & +1 & 0  & +1 & 0 & 0 & +1 \\
B & 0 & +$\frac{1}{3}$  & 0  &+$\frac{1}{3}$ & +$\frac{1}{3}$ & 0 \\
 \end{tabular}
 \end{center}
%
where, for once, we have bowed to tradition and employed the non-integer normalisation of the baryon current. 
Both $L$ and $B$ suffer mixed anomalies with both $SU(2)$ and $U(1)_{\tilde{Y}}$. They are
%
%
%
\be \sum L\,SU(2)^2 = \sum B\,SU(2)^2  = -1 \ \ {\rm and} \ \ \sum L\,\tilde{Y}^2 = \sum B\,\tilde{Y}^2 = +18 \nn\ee
We recover the well known fact that the combination $B-L$ is non-anomalous. Meanwhile, under a transformation of $L$, parameterised by $\alpha_L$, we have
\be L: \ \theta_2\rightarrow \theta_2-\alpha_L\ \ ,\ \ \tilde{\theta}\rightarrow \tilde{\theta} + 18\alpha_L\nn\ee
The linear combination $\tilde{\theta} + 18\theta_2$ is physical and cannot be rotated away. We will see the  interpretation of this shortly.

\subsection{Electroweak Symmetry Breaking}

Let's now see what becomes of our line operators after electroweak symmetry breaking. The Higgs field $H$ lies in the $({\bf 2},{\bf 1})_3$  representation and condenses, breaking $U(1)_{\tilde{Y}}\times SU(2)\rightarrow U(1)_{em}$ of electromagnetism. We will denote the electric charges of $U(1)_{em}$ as $Q$ and the magnetic charges as $G$. We choose to normalise electric charges such that the electron has $Q=-1$. 

\para
The allowed electric and magnetic charges under $U(1)_{em}$ depend on our choice of discrete quotient $\Gamma$.  The electric charge under $U(1)_{em}$ is given by
\be Q =  \frac{q}{6} + \lambda_2^e\nn\ee
with $q$ the $U(1)_{\tilde{Y}}$ charge and with the normalisation $\lambda_2^2\in {\bf Z}$ so that, for example, $\lambda_2^e = \pm 1$ corresponds to the fundamental representation of $SU(2)$. (Written in terms of $Y=\tilde{Y}/6$, this takes the more familiar form $Q = Y + \lambda^e_2$.)

\para
Meanwhile, after condensation  of the Higgs, most 't Hooft and dyonic line operators with  $U(1)_{\tilde{Y}}$ or $SU(2)$ magnetic charge exhibit an area law. Those that remain deconfined obey the condition
\be 6g = \lambda_2^m\ \ \ \Rightarrow\ \ \  6g = z_2^m\ {\rm mod}\ 2\nn\ee
The resulting magnetic charge under $U(1)_{em}$ is given by
\be G=6g\nn\ee
 In this normalisation, the Dirac quantisation condition for pure electromagnetism reads $QG\in {\bf Z}$. 
We can now describe the  spectrum of electric and magnetic $U(1)_{em}$ charges for each quotient. 

\para
\underline{$\Gamma= {\bf 1}$:}\ \ The minimum electric charge is $Q=1/6$ and the minimum magnetic charge is $G=6$. These arise, for example, from the Wilson line $({\bf 1},{\bf 1})_1$ and the 't Hooft line with $g=1$ and $\lambda_2^m = 6$. 
\para
\underline{$\Gamma= {\bf Z}_2$:}\ \ The minimum electric charge is $Q=1/3$ and the minimum magnetic charge is $G=3$. These arise, for example, from the Wilson line $({\bf 1},{\bf 1})_2$ and the 't Hooft line with $g=1/2$ and $\lambda_2^m = 3$. 
\para
\underline{$\Gamma= {\bf Z}_3$:}\ \  The minimum electric charge is $Q=1/6$ and the minimum magnetic charge is $G=2$. These arise, for example, from the Wilson line $({\bf 1},{\bf 3})_1$ and the 't Hooft line with $g=1/3$ and $\lambda_2^m = 2$. 
\para
\underline{$\Gamma= {\bf Z}_6$:}\ \ The minimum electric charge is $Q=1/3$ and the minimum magnetic charge is $G=1$. These arise, for example, from the Wilson line $({\bf 1},{\bf 3})_{-2}$ and the 't Hooft line with $g=1/6$ and $\lambda_2^m = 1$.

\para
Note that for $\Gamma= {\bf Z}_3$ and $\Gamma  = {\bf Z}_6$, the spectrum is not consistent with the naive,  electromagnetic Dirac quantisation $QG\in {\bf Z}$. This is simply the statement that a minimum Dirac monopole is inconsistent with the fractional charge of the quarks. The resolution to this was given long ago  \cite{corrigan}: the magnetic monopole must also carry colour magnetic charge,  and this  provides an extra contribution to the Dirac quantisation condition, rendering the spectrum consistent. (This fact is also emphasised in \cite{preskill} in the context of GUT monopoles.) Indeed, it is simple to check from  Figures 4 and 5 that the relevant 't Hooft lines do indeed carry $SU(3)$ magnetic charge. This is tantamount to the fact that, in these cases, the low-energy gauge group is actually $U(3)$ rather than $U(1) \times SU(3)$.

\para
We denote the electromagnetic field strength as $F$, again normalised such that the electron carries charge $-1$. After symmetry breaking, the electromagnetic theta term is
\be S_\theta = \frac{\theta_{\tilde{Y}}+ 18\theta_2}{64 \pi^2} \int d^4x\ {}^\star F F\nn\ee
%
%
%
We see that  $\theta_{em} =(\theta_{\tilde{Y}}+ 18\theta_2)/4$,  precisely the combination that cannot be removed by a chiral rotation. This, of course, is no coincidence: it follows from 't Hooft anomaly matching and the fact that the  Higgs multiplet $({\bf 2},{\bf 1})_3$ can give mass to all chiral fermions, leaving behind the vector-like theory of QED. 

\para
The range of $\theta_{em}$ depends on the quotient $\Gamma$. For $\Gamma = {\bf 1}, {\bf Z}_2$, the gauge group is $U(1) \times SU(3)$ and $\theta_{em} \in [0,2\pi Q^2)$ where $Q$ is the minimum charge. For $\Gamma = {\bf Z}_3, {\bf Z}_6$, the gauge group is $U(3)$ and $\theta_{em} \in [0,18\pi Q^2)$. In particular, when $\Gamma = {\bf Z}_6$ we have $\theta_{em} \in [0,2\pi)$ despite the presence of fractional quark charge.

\section{Summary}

The global structure of the Standard Model gauge group depends on the choice of quotient $\Gamma$. The differences in these theories described above are rather formal in nature. Just because a line operator exists in a theory does not mean that it is available to experimenters.  It is clearly interesting to better understand the physical implications of the different choices of $\Gamma$ to see if they may reveal themselves in some way in our world. 

\para
Some minor, and fairly cheap, respite can be found  in the conjecture that, in any theory with  gravity,  the full set of  charges carried by  line operators are also carried by  dynamical objects \cite{joe}. The best arguments for this come from black hole physics \cite{banks}. In the absence of gravity, one can always decouple fields by taking their mass to infinity, leaving behind only the non-dynamical line operators. In a theory with gravity, this is not possible: the backreaction of the line operator will eventually form a black hole, which carries the appropriate electric and magnetic charges. 

\para
Conversely, the global structure of the gauge group determines the fluxes that are allowed through cycles in a non-trivial spacetime. A black hole provides such a spacetime, with an ${\bf S}^2$ horizon, and the fluxes which can thread this are determined by the choice of $\Gamma$. 

\para
The further requirement that elementary particles should not form black holes \cite{weakest} suggests that there are new magnetic (and possibly electric) particles to be found whose charges depend on $\Gamma$. Obviously,  if a neutral quark is discovered, transforming in the  $({\bf 1},{\bf 3})_0$ representation of $G$, then  we must take $\Gamma =  {\bf Z}_2$ or ${\bf 1}$. In contrast, as we saw above, the discovery of a magnetic monopole, consistent with the minimum Dirac quantisation with respect to the electron, but not with respect to the quark,  would mean that $\Gamma = {\bf Z}_6$. 
 Of course, these predictions are rather toothless: the particles have a mass which is  constrained only by the Planck scale  are unlikely to be abundant; one must hope that Nature is kind \cite{moedal}.

\para
One can ask if there are more subtle ways to distinguish between the theories. As explained in \cite{ast}, and reviewed in Section \ref{reviewsec}, after confinement the Yang-Mills theory with gauge group $SU(N)/{\bf Z}_N$ exhibits topological order, with an emergent ${\bf Z}_N$ magnetic gauge symmetry. This gives rise to the possibility of more interesting physics arising  in spacetimes with non-trivial topologies or boundaries. For the Standard Model, there exists a magnetic ${\bf Z}_3$ symmetry arising when $\Gamma = {\bf Z}_3$ or ${\bf Z}_6$ but the states which carry charge under this also carry magnetic charge under under $U(1)_{em}$. It would be interesting to see if this has any implications for physics in the presence of non-trivial topology or, more interestingly, in dynamical spacetime.  This may allow us to answer the basic  question: what is the gauge group of the Standard Model?

\subsection*{Acknowledgements}

My thanks to Bobby Acharya, Nick Dorey, Zohar Komargodski, Tony Padilla and John Terning for useful discussions. 
 This work was supported by STFC grant ST/L000385/1 and by the European Research Council under
the European Union's Seventh Framework Programme (FP7/2007-2013), ERC grant
agreement STG 279943, ÒStrongly Coupled SystemsÓ. I am a  Wolfson Royal Society Research Merit Award holder.

\end{document}